\documentclass[letter,aps,prx,article,twocolumn,floatfix,superscriptaddress]{revtex4-2}
\date{\today}
\usepackage{epsfig}
\usepackage{graphics}
\usepackage{dcolumn}
\usepackage{bm}
\usepackage[colorlinks,citecolor = blue,linkcolor=blue,hyperindex,CJKbookmarks]{hyperref}
\usepackage{float}
\usepackage{hyperref}
\usepackage{comment}
\usepackage{mathtools}
\usepackage{accents}

\newcommand{\eph}{{\it e}-ph}

\begin{document}

\title{Emergent $s+id$ Superconductivity from the Interplay between Electronic Correlations and Electron‑Phonon Coupling in $\mathrm{R}_{1-x}\mathrm{Sr}_x\mathrm{NiO}_2$}

\author{Zi Yuan}
\affiliation{Department of Physics, Northeastern University, Shenyang 110819, China}

\author{Jun Zhan}
\affiliation{Beijing National Laboratory for Condensed Matter Physics and Institute of Physics, Chinese Academy of Sciences, Beijing 100190, China}
\affiliation{School of Physical Sciences, University of Chinese Academy of Sciences, Beijing 100190, China}

\author{Xianxin Wu}
\affiliation{CAS Key Laboratory of Theoretical Physics, Institute of Theoretical Physics, Chinese Academy of Sciences, Beijing 100190, China}

\author{Shaozhi Li}
\email[Corresponding author: ]{lishaozhi@mail.neu.edu.cn}
\affiliation{Department of Physics, Northeastern University, Shenyang 110819, China}

\date{\today}
\begin{abstract}
Recent tunneling measurements on infinite-layer nickelates have revealed spatially varying superconducting symmetries, whose microscopic origin remains unclear. Motivated by this observation, we investigate the interplay between electron correlations and electron-phonon interactions in infinite-layer nickelates by combining first-principles calculations with the fluctuation-exchange–Migdal-Eliashberg theory. Our calculations show that spin fluctuations yield robust $d$-wave superconductivity on the Ni $d_{x^2-y^2}$ orbital, whereas electron-phonon coupling induces $s$-wave pairing on an interstitial orbital, leading to an $s+id$ superconducting state. The emergence of the $s$-wave component is strongly carrier-density dependent: an intermediate electron-phonon coupling of $\lambda=0.4$ stabilizes the $s+id$ state at $n=0.9$ but not at $n=0.8$. These results imply that local oxygen defects tune the local electron density and form finite-size domains with distinct pairing symmetries, offering a compelling explanation for the spatially inhomogeneous superconducting symmetries observed in experiments.
\end{abstract}

\maketitle

In the pursuit of new high-temperature superconductors~\cite{DopingLee2006,CorrelatedDagotto1994,HighPaglione2010,HighSi2016}, nickelate superconductors have attracted considerable attention~\cite{SuperconductingKreisel2022, SignaturesSun2023, High-temperatureZhang2024, VisualizationDong2024, BulkWang2024, ElectronicLiu2024, Orbital-dependentYang2024, EvidenceChen2024, ElectronicChen2024, IdentificationLi2025, BulkLi2026, RecentWang2025, StabilizationLu2026, SignaturesKo2025, Ambient-pressureZhou2025, Fermi-liquidHsu2026, SuperconductivityLiu2025, Angle-resolvedLi2025, ElectronicWang2025, SuperconductivityZhou2026, ReducingTarn2026, EnhancedLi2026, SuperconductivityHao2025, SuperconductingWang2026, NodelessShen2026, BilayerLuo2023, CorrelatedChristiansson2023, InterlayerLu2024, BilayerQu2024, SuperconductivitySchlomer2024, StrongZhang2024, HighLuo2024, PossibleSakakibara2024, PressureJiang2024, HighlyWang2025, CorrelatedYue2025, TheoryJiang2025, CooperationZhan2025, FermiWang2026,BulkChow2025,EnhancedYang2025,SuperconductivitySahib2025,CriticalJiang2020,ElectronicSun2025,Cuprate-likeDing2024,Linear-in-temperatureLee2023,NickelateOsada2021,ARossi2022,AGu2020,ChargePeng2021,Core-LevelHigashi2021,ElectronicHepting2020,ElectronicBeen2021,Infinite-layerKang2023,Many-BodyKarp2020,ModelSakakibara2020,MultiorbitalLechermann2020,NickelateKitatani2020,orbitalAdhikary2020,Pressure-inducedWang2022,ScreeningPlienbumrung2022,SimilaritiesBotana2020,SuperconductingLi2020,SuperconductivityXiao2024,Superconductivityzeng2022,SuperconductivityParzyck2025,SuperconductivityLuo2023,Two-bandHu2019,Type-IIZhang2020,PreemptedMeier2024}. Superconductivity in nickelates was first reported in the hole-doped infinite-layer compound $\mathrm{Nd}_{0.8}\mathrm{Sr}_{0.2}\mathrm{NiO}_2$, with a superconducting (SC) transition temperature ($T_c$) of approximately 9--15 K~\cite{SuperconductivityLi2019}. A recent ARPES measurement on $\mathrm{La}_{0.8}\mathrm{Sr}_{0.2}\mathrm{NiO}_2$ reveals a large Fermi surface from the Ni $d_{x^2-y^2}$ orbital and a small 3D electron pocket at the Brillouin zone corner from an uncorrelated conduction band~\cite{ElectronicSun2025}.
The Ni $d_{x^2-y^2}$ band shows strong band renormalization, while the uncorrelated band exhibits negligible renormalization.
DFT calculations confirm these bands are well described by a two-orbital model including the Ni $d_{x^2-y^2}$ orbital and an interstitial orbital (IO) between Ni layers~\cite{orbitalAdhikary2020,AGu2020,ScreeningPlienbumrung2022}.



Superconductivity in infinite-layer (IL) nickelates was initially attributed to the Ni $d_{x^2-y^2}$ band with $d$-wave pairing~\cite{ModelSakakibara2020,RobustWu2020,SuperconductivityYusuke2022}. However, this conventional scenario has been challenged by recent experimental advances. For instance, scanning tunneling spectroscopy measurements on $\mathrm{Nd}_{1-x}\mathrm{Sr}_x\mathrm{NiO}_2$ thin films have revealed two distinct superconducting (SC) gap features corresponding to $s$-wave and $d$-wave pairing symmetries~\cite{SingleGu2020}. The nature of the SC gap is spatially dependent on the probing tip position: pure $s$-wave or pure $d$-wave gaps dominate in certain regions, while coexisting $s$- and $d$-wave gaps are observed in others. Complementarily, London penetration depth measurements further corroborate an $s+id$-wave superconducting gap in Nd-based IL nickelate systems~\cite{PairingChow2023}. Collectively, these observations motivate us to re-examine the microscopic origin of superconductivity in IL nickelates, moving beyond the conventional \(d\)-wave paradigm.


An early study showed that the $s+id$ SC state in IL nickelates could originate from the doping effect based on a two-orbital $t$-$J$ model~\cite{DistinctWang2020}. In contrast, a recent DFT+GW study showed that the electron-phonon ({\eph}) coupling strength in \(\mathrm{Nd}_{0.8}\mathrm{Sr}_{0.2}\mathrm{NiO}_2\) is strong enough to induce a SC transition temperature of 25~K at optimal hole doping. Moreover, it showed that the two distinct types of SC gaps could arise from different pairing strengths on the Ni \(d\) and Nd \(d\) orbitals~\cite{TwogapLi2024}. Nevertheless, this DFT+GW framework excludes strong on-site Coulomb interactions. As a result, this computational scheme could not fully account for the dual-gap SC behavior observed in $\mathrm{Nd}_{1-x}\mathrm{Sr}_x\mathrm{NiO}_2$.
Even so, the findings from Ref.~\cite{TwogapLi2024} prompt us to revisit the contribution of phonons to the SC pairing mechanism in IL nickelates.

Our work investigates the interplay between electron–electron and {\eph} interactions in shaping superconductivity in IL nickelates. We develop a two-orbital model for doped $\mathrm{La}_{1-x}\mathrm{Sr}_x\mathrm{NiO}_2$ that incorporates both on-site Coulomb repulsion on the Ni $d_{x^2-y^2}$ orbital and coupling between the IO and oxygen $A_{1g}$ phonon mode. Employing the fluctuation-exchange plus Migdal–Eliashberg (FLEX–ME) approach, we show that robust $d$-wave superconductivity arises from the $d_{x^2-y^2}$ orbital across a wide doping range. Notably, an intermediate {\eph} coupling (i.e. $\lambda\approx 0.4$) induces an $s$-wave SC gap on the IO at $n\approx 0.9$, yielding a mixed $s+id$-wave pairing state, while such coupling is too weak to stabilize $s$-wave superconductivity at $n=0.8$. These results indicate that the spatially dependent pairing symmetry observed experimentally in $\mathrm{Nd}_{1-x}\mathrm{Sr}_x\mathrm{NiO}_2$ could originate from local oxygen defects, which modulate local carrier density on the nanoscale. Therefore, our work offers a microscopic physical origin for the positional variation of SC symmetries observed in tunneling spectroscopy.

\begin{figure}[t]
\begin{center}
\includegraphics[width=0.99\columnwidth]{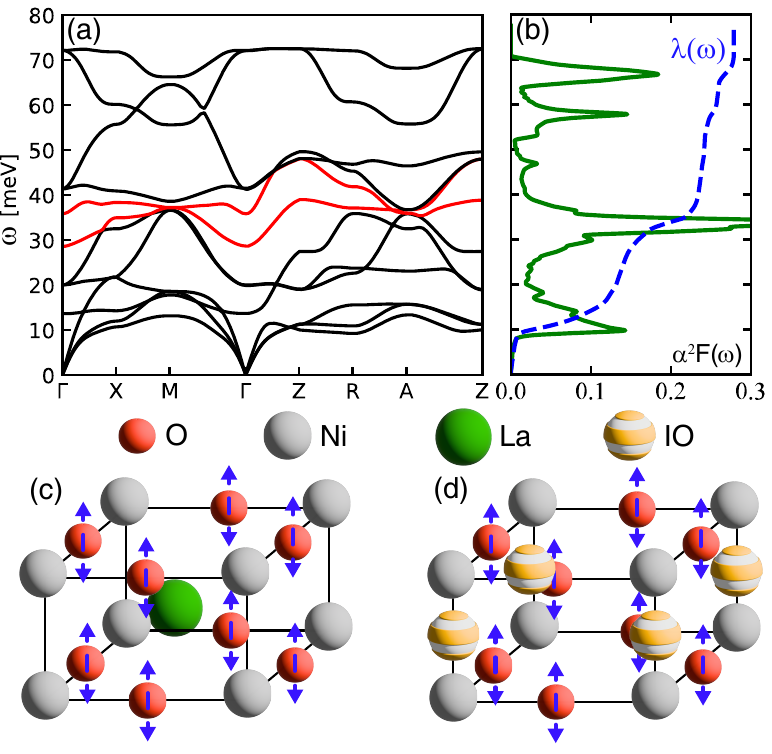}
\caption{
\label{fig:1}Phonon properties of $\mathrm{La}\mathrm{NiO}_2$. (a) Phonon dispersion. (b) Electron‑phonon coupling function $\alpha^2 F(\omega)$. (c) $A_{1g}$ vibration mode of the in‑plane oxygen atoms. (d) $A_{1g}$ vibration mode for the effective two‑orbital model. IO stands for interstitial orbital.
}
\end{center}
\end{figure}

We first investigate the phonon properties of $\mathrm{LaNiO}_2$ using the Quantum ESPRESSO package with the norm-conserving Vanderbilt (ONCV) pseudopotential library~\cite{QUANTUMgIANNOZZI2009,OptimizedHamann2013}.
Figure~\ref{fig:1}(a) presents the phonon band structure calculated along a high-symmetry path in the Brillouin zone.
The red curve denotes the $A_{1g}$ phonon mode associated with in-plane oxygen vibrations; the corresponding atomic displacement pattern is illustrated in Fig.~\ref{fig:1}(c).
Figure~\ref{fig:1}(b) displays the {\eph} spectral function $\alpha^2F(\omega)$ as a function of energy $\omega$.
A prominent peak appears in $\alpha^2F(\omega)$ within the energy range of the $A_{1g}$ phonon (30--40~meV), signifying strong {\eph} coupling for this mode.
A similar feature has also been reported in the bilayer nickelate superconductor $\mathrm{La}_3\mathrm{Ni}_2\mathrm{O}_7$~\cite{CooperationZhan2025}.
From the total spectral function, we extract a dimensionless {\eph} coupling strength $\lambda \approx 0.3$, of which the $A_{1g}$ phonon contributes roughly one-third ($\lambda_{A_{1g}} \approx 0.1$).

Our DFT calculations do not include on-site Coulomb interactions (DFT+U) or higher-order corrections from quantum dynamics~\cite{TwogapLi2024}. However, including electronic correlations has been shown to enhance the {\eph} coupling~\cite{CorrelationYin2013}. Moreover, Ref.~\cite{TwogapLi2024} demonstrates that incorporating GW corrections into DFT can increase the {\eph} coupling strength by a factor of {\it five} in $\mathrm{Nd}_{0.8}\mathrm{Sr}_{0.2}\mathrm{NiO}_2$. Together, these results indicate that DFT-based calculations typically underestimate the true {\eph} coupling in strongly correlated materials.
The value obtained in Fig.~\ref{fig:1} should therefore be regarded as a lower bound.
Accordingly, we estimate that the actual {\eph} coupling strength for the $A_{1g}$ phonon mode in $\mathrm{LaNiO}_2$ lies in the range of $0.1$--$0.5$.

Our study focuses on the oxygen $A_{1g}$ phonon mode. Lower-frequency phonon modes produce a very low SC transition temperature, whereas higher-frequency modes feature excessively weak {\eph} coupling.
To examine SC pairing, we employ the two-orbital model originally introduced in Ref.~\cite{ElectronicSun2025}, comprising the Ni $d_{x^2-y^2}$ orbital and an interstitial $s$-like orbital; the orbital layout is sketched in Fig.~\ref{fig:1}(d).
The full Hamiltonian takes the form
\begin{align}
H &= \sum_{{\bf k},\sigma} \Psi_{{\bf k},\sigma}^{\dagger} H_0({\bf k}) \Psi_{{\bf k},\sigma} + U_d\sum_{{\bf r}} n_{{\bf r},d,\uparrow}n_{{\bf r},d,\downarrow} \notag \\
& + \sum_{{\bf q},\gamma}\Omega b^{\dagger}_{{\bf q},\gamma} b_{{\bf q},\gamma} + H_\mathrm{eph},
\end{align}
where the two-component spinor $\Psi_{{\bf k},\sigma} = (d_{{\bf k},\sigma}, s_{{\bf k},\sigma})^{\mathsf{T}}$ denotes annihilation operators for the Ni $d_{x^2-y^2}$ and effective $s$ orbitals at momentum ${\bf k}$ and spin $\sigma$.
The operator $n_{\mathbf{r}, d, \sigma}$ gives the occupation number for the Ni $d_{x^2-y^2}$ orbital at lattice site $\mathbf{r}$, and $U_d$ denotes the corresponding on-site Coulomb interaction parameter.
Unless noted otherwise, we fix $U_d = 4$~eV for all subsequent calculations.
We omit on-site Coulomb repulsion for the $s$ orbital because ARPES measurements show negligible band renormalization for the associated band~\cite{ElectronicSun2025}.
The kinetic tight-binding term $H_0(\mathbf{k})$ accounts for intraorbital and interorbital hopping matrix elements, with its explicit form provided in the Supplemental Material~\cite{SM}.
The parameter $\Omega$ denotes the energy of the optical $A_{1g}$ phonon, set to $35$~meV; $b_{{\bf q},\gamma}$ is the phonon annihilation operator at momentum ${\bf q}$, and the index $\gamma=x,y$ labels the two oxygen atoms within one unit cell. The electron-phonon interaction is defined as
\begin{align}
H_\mathrm{eph} &= g\sum_{\mathbf{r},\sigma} n_{\mathbf{r},s,\sigma} \bigl( \hat{u}_{\mathbf{r},x}+\hat{u}_{\mathbf{r},y}+\hat{u}_{\mathbf{r}-\mathbf{a},x}+\hat{u}_{\mathbf{r}-\mathbf{b},y} \notag \\
& -\hat{u}_{\mathbf{r}+\mathbf{c},x}-\hat{u}_{\mathbf{r}+\mathbf{c},y}-\hat{u}_{\mathbf{r}-\mathbf{a}+\mathbf{c},x}-\hat{u}_{\mathbf{r}-\mathbf{b}+\mathbf{c},y} \bigr),
\end{align}
in which $\hat{u}_{\mathbf{r},\gamma}$ stands for the ionic displacement operator.
We neglect direct coupling between Ni $d_{x^2-y^2}$ electrons and the $A_{1g}$ phonons, as this coupling enters only at nonlinear {\eph} order~\cite{QuasiparticleLi2015}.
For small-amplitude lattice displacements, such nonlinear contributions are far weaker than the linear first-order coupling retained here.
The dimensionless {\eph} coupling strength is defined as $\lambda = \frac{N_F g^2}{M\Omega^2}$, where $N_F$ is the Fermi-level density of states evaluated from the bare Hamiltonian $H_0$, $g$ denotes the {\eph} coupling strength, and $M$ is the mass of the oxygen ion.

\begin{figure}[t]
\begin{center}
\includegraphics[width=0.99\columnwidth]{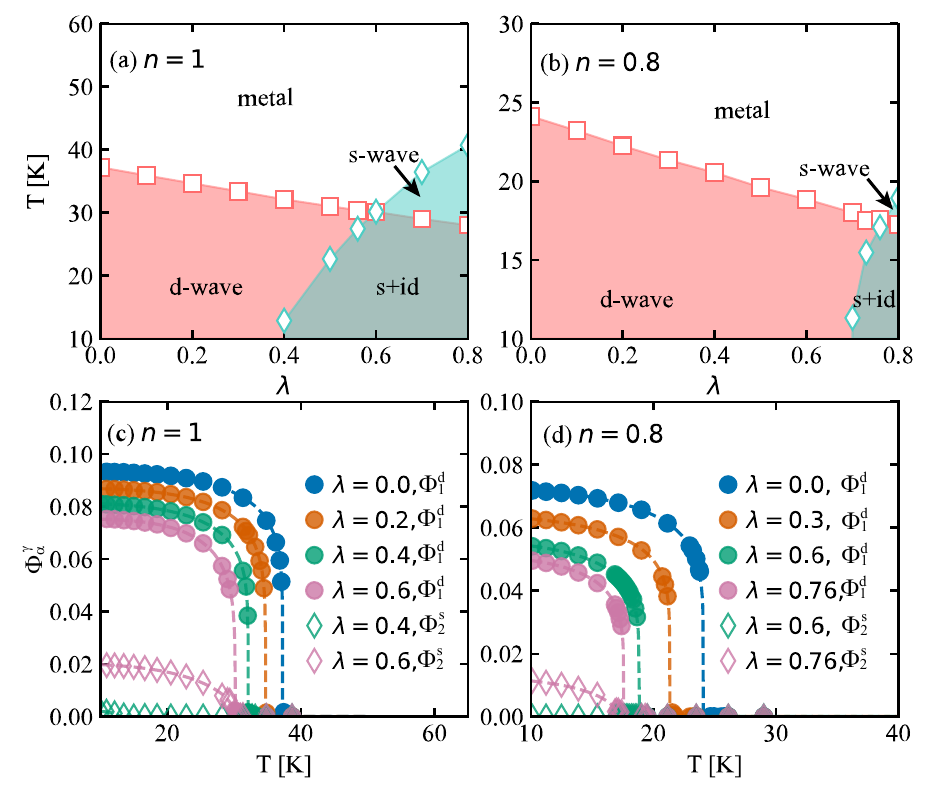}
\caption{(a, b) Phase diagrams in the plane of temperature $T$ and electron–phonon coupling $\lambda$ for electron densities $n = 1$ and $n = 0.8$, respectively. (c, d) The averaged superconducting order parameter $\Phi_{\gamma}^{d(s)}$ as a function of temperature for various values of $\lambda$ at $n = 1$ and $n = 0.8$, respectively. 
\label{fig:2} 
}
\end{center}
\end{figure}

We employ the combined fluctuation-exchange (FLEX) and Migdal–Eliashberg (ME) formalism to map out the phase diagram of our three-dimensional two-orbital model~\cite{EnhancedRademaker2021}.
This approach simultaneously captures $d$-wave superconductivity driven by spin fluctuations and $s$-wave superconductivity arising from charge fluctuations.
Full computational details are documented in the Supplemental Material~\cite{SM}.
All numerical simulations are carried out on a $22\times22\times22$ lattice unless specified otherwise.

Figures.~\ref{fig:2}(a) and~\ref{fig:2}(b) display the phase diagrams for electron densities $n = 1$ and $n = 0.8$, corresponding to $x = 0$ and $x = 0.2$ in $\mathrm{La}_{1-x}\mathrm{Sr}_x\mathrm{NiO}_2$, respectively. The transition temperature is extracted from the temperature dependence of the anomalous self-energy $\Phi_{\alpha}(\mathbf{k}, i\omega_n)$ evaluated at the lowest Matsubara frequency. Specifically, we project $\Phi_{1}(\mathbf{k}, i\omega_0)$ of orbital 1 ($d$ orbital) onto the basis $(\cos k_x - \cos k_y)$, and $\Phi_{2}(\mathbf{k}, i\omega_0)$ of orbital 2 (IO orbital) onto a constant basis, yielding the momentum-averaged $d$-wave and $s$-wave pairing order parameters $\Phi_{1}^d$ and $\Phi_{2}^s$, respectively. We subsequently fit the temperature dependence of $\Phi_{1}^d$ and $\Phi_{2}^s$ with the empirical forms $y = a\left[1-(T/T_c)^b\right]^c$ and $y = a\sqrt{1-(T/T_c)^b}$, respectively.
The dashed lines plotted in Figs.~\ref{fig:2}(c) and~\ref{fig:2}(d) represent these fitting outcomes.

As shown in Fig.~\ref{fig:2}(c), at $n = 1$ and $\lambda = 0$, the $d$-wave SC transition temperature is approximately $T_{d,c} = 39$ K. This critical temperature decreases linearly with increasing $\lambda$. For $\lambda > 0.4$, a coexistence regime of $s$-wave and $d$-wave SC states emerges. When $\lambda > 0.6$, the $s$-wave transition temperature $T_{s,c}$ exceeds $T_{d,c}$; for $0.4 < \lambda < 0.6$, $T_{s,c}$ is lower than $T_{d,c}$. In comparison to the phase diagram at $n=1$, a larger {\eph} coupling strength is needed to stabilize the $s$-wave SC state at $n=0.8$. This behavior originates from the reduced electron density residing on orbital 2 at $n=0.8$ relative to the $n=1$ case.

Finite-size corrections to these SC transition temperatures are quantified in the Supplemental Material~\cite{SM}.
For $\lambda=0.8$ and $n=1$, $T_{s,c}$ falls gradually from $47$~K on an $18\times18$ lattice to $41$~K for $22\time22$, and further down to $37$~K at $26\times26$.
Extrapolating these data to the thermodynamic limit gives $T_{s,c}=13.7$~K.
By contrast, the $d$-wave critical temperature $T_{d,c}\approx28$~K exhibits negligible lattice-size dependence. This outcome demonstrates that, in the thermodynamic limit at $\lambda=0.8$, $T_{s,c}$ is in fact smaller than $T_{d,c}$.
We further extrapolate $T_{s,c}$ to the thermodynamic limit across a range of $\lambda$ and identify a threshold coupling $\lambda\approx0.7$, above which finite $T_{s,c}>0$ emerges.
For $\mathrm{La}_{1-x}\mathrm{Sr}_x\mathrm{NiO}_2$, the physical {\eph} coupling strength is expected to lie below this critical threshold, meaning bulk $s$-wave superconductivity cannot be sustained in the stoichiometric crystal. Nevertheless, oxygen defects can create nanoscale finite domains, where $s$-wave SC signals may still be detectable experimentally.

\begin{figure}[t]
\begin{center}
\includegraphics[width=0.99\columnwidth]{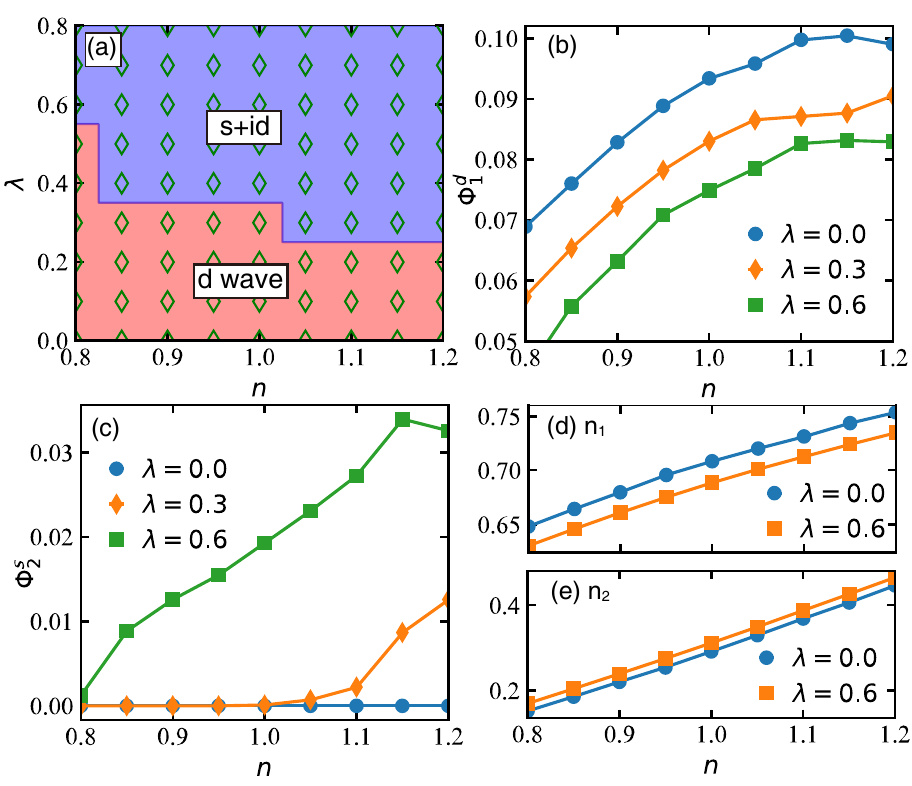}
\caption{(a) Phase diagram in the $(n,\lambda)$ plane at $T=16$ K.
(b,c) Averaged $d$- and $s$-wave superconducting order parameters, $\Phi_{1}^{d}$ and $\Phi_{2}^{s}$, as functions of $n$.
(d,e) Orbital-resolved electron densities for the $d$ and $s$ orbitals.
\label{fig:3} 
}
\end{center}
\end{figure}

\begin{figure}[t]
\begin{center}
\includegraphics[width=0.99\columnwidth]{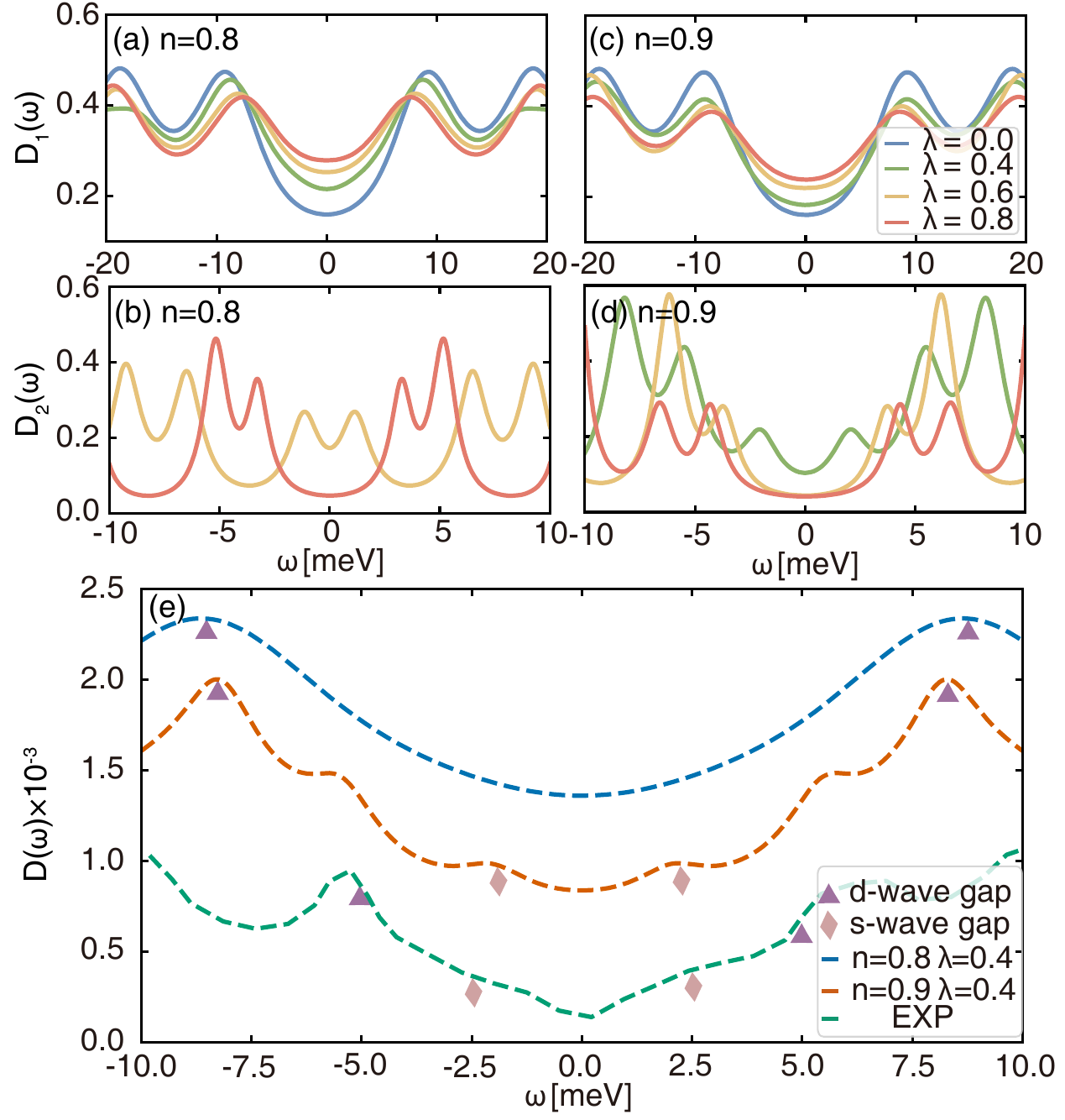}
\caption{ (a,c) Density of states $D_1(\omega)$ for orbital 1 at fillings $n=0.8$ and $n=0.9$, respectively. (b,d) Density of states $D_2(\omega)$ for orbital 2 at fillings $n=0.8$ and $n=0.9$, respectively. (e) Total density of states $D(\omega)$. The green dashed curve represents the single-particle tunneling spectrum of $\mathrm{Nd}_{0.8}\mathrm{Sr}_{0.2}\mathrm{NiO}_2$ from Ref.~\cite{SingleGu2020}.
\label{fig:4} 
}
\end{center}
\end{figure}

To further elucidate the density dependence of superconductivity, we compute the phase diagram in the $(n,\lambda)$ plane at $T=16$~K, as shown in Fig.~\ref{fig:3}(a). Green diamonds mark the parameter points where calculations were performed. Figures.~\ref{fig:3}(b) and~\ref{fig:3}(c) display the averaged SC order parameters $\Phi_{1}^{d}$ and $\Phi_{2}^{s}$ for $\lambda=0$, $0.3$, and $0.6$, respectively. At $n=0.8$, the critical {\eph} coupling strength for the onset of the $s+id$ state is $\lambda_c \approx 0.55$ (with $\Phi_{2}^{s} \approx 0.0012$ at $\lambda=0.6$). Notably, $\lambda_c$ decreases rapidly with increasing electron density: it is about $0.35$ at $n=1.0$ and falls further to $\sim 0.25$ at $n=1.1$. 
These results imply that IL nickelates exhibit substantial particle–hole asymmetry, and that the $s+id$ SC state is far easier to stabilize at elevated electron doping levels~\cite{ElectronEzra2026}.

Our two-orbital model predicts a robust $d$-wave SC phase persisting over a broad range of electron densities. This finding contrasts with early experimental measurements that identified a SC dome solely within the narrow density window $0.7 < n < 0.85$~\cite{Superconductivityzeng2022}. However, recent experiments have confirmed superconductivity in the parent IL nickelate $\mathrm{NdNiO}_2$~\cite{SuperconductivityParzyck2025}. The absence of superconductivity in the parent compound reported in Ref.~\cite{Superconductivityzeng2022} is therefore unlikely to be intrinsic, but instead may originate from the disorder effect~\cite{SuperconductivityParzyck2025}.

Next, we examine the impact of the {\eph} coupling on the orbital-resolved occupations. As shown in Figs.~\ref{fig:3}(d) and~\ref{fig:3}(e), coupling to the $A_{1g}$ phonon suppresses the electron occupation on orbital 1 while slightly enhancing that on orbital 2. This charge redistribution may underlie the reduction of $T_{d,c}$ observed in Fig.~\ref{fig:2}(a).  Moreover, the enhanced charge fluctuation on orbital 1, arising from the proximity effect, may further contribute to this reduction.

To extract the SC gap, we perform analytic continuation of the local Green's function using the Pad\'e approximation on a $24\times24\times24$ lattice at $T=16$~K. The frequency broadening factors in the  Pad\'e approximation are set as $\delta_1=0.004$ for orbital 1 and $\delta_2=0.0008$ for orbital 2, respectively. The smaller broadening factor for orbital 2 is adopted due to the smaller $s$-wave SC gap compared to the $d$-wave SC gap. Figures~\ref{fig:4}(a) and~\ref{fig:4}(b) show the local density of states (DOS) $D_{1}(\omega)$ for orbital 1 at $n=0.8$ and $n=0.9$, respectively, for different values of $\lambda$. The $d$-wave SC gap slightly decreases with increasing $\lambda$; for example, at $n=0.8$, it is about $9$~meV at $\lambda=0$ and reduces to $\sim 7.5$~meV at $\lambda=0.8$. Figures~\ref{fig:4}(c) and~\ref{fig:4}(d) display $D_{2}(\omega)$ for orbital 2 at $n=0.8$ and $n=0.9$, respectively. In contrast, the $s$-wave SC gap increases significantly with $\lambda$: at $n=0.8$, the SC gap is about 1~meV at $\lambda=0.6$, where the system lies close to the $s+id$ phase boundary, and grows to $\sim 3.2$~meV at $\lambda=0.8$.

The $s+id$ SC state has been observed in single-particle tunneling measurements on $\mathrm{Nd}_{0.8}\mathrm{Sr}_{0.2}\mathrm{NiO}_2$~\cite{SingleGu2020}, as shown by the green dashed curve in Fig.~\ref{fig:4}(e). The extracted $s$-wave and $d$-wave gaps are approximately $2.5$~meV and $5$~meV, respectively, marked by the diamond and triangle symbols. Notably, the pairing symmetry in $\mathrm{Nd}_{0.8}\mathrm{Sr}_{0.2}\mathrm{NiO}_2$ reported in Ref.~\cite{SingleGu2020} depends on the tip position: in some regions, a $d$-wave gap is observed, while in others, an $s$-wave gap appears. We attribute this position dependence to oxygen defects, which locally modify the electron density. 

\begin{figure}[t]
\begin{center}
\includegraphics[width=0.95\columnwidth]{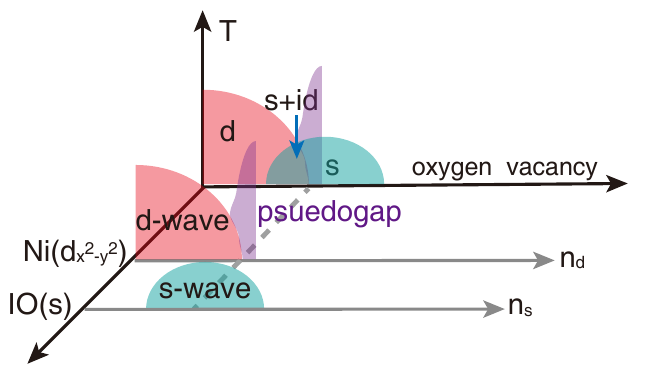}
\caption{ 
\label{fig:5} Schematic illustration of the effect of oxygen vacancies on superconductivity in $\mathrm{R}_{0.8}\mathrm{Sr}_{0.2}\mathrm{NiO}_2$ ($R = \mathrm{Nd}, \mathrm{La}$) as a function of vacancy concentration. Here, $n_d$ and $n_s$ denote the electron densities on the Ni $d_{x^2-y^2}$ and IO orbitals, respectively. 
}
\end{center}
\end{figure}

To further support this interpretation, we compare our calculated DOS with the experimental tunneling spectrum shown in Fig.~\ref{fig:4}(e).
To suppress the strong $s$-wave coherence peaks induced by the small value of $\delta$, we approximate the total density of states as a weighted sum $\delta_1 D(\omega) + \delta_2 D(\omega)$ (see Supplemental Material for details~\cite{SM}).
For $U_d = 4$~eV, $\lambda = 0.4$, and $n = 0.9$, the calculated $s$-wave SC gap is approximately $2.3$~meV, in good agreement with the experimental measurement.
In contrast, $s$-wave superconductivity is found to be completely suppressed at $n = 0.8$.
For the $d$-wave gap, the calculated value is roughly $7.5$~meV for both $n = 0.8$ and $n = 0.9$, slightly exceeding the experimental value.
Notably, the Hubbard $U_d$ parameter used here ($U_d = 4$~eV) falls within the widely accepted range of $3$--$4$~eV for nickelate superconductors, consistent with previous studies~\cite{MagneticKlett2022,ElectronicSun2025}.
Therefore, the slight overestimation of the $d$-wave gap can be attributed to the inherent limitations of the FLEX approximation~\cite{HubbardDong2022}.

Figure~\ref{fig:5} summarizes the effects of oxygen vacancies on superconductivity in $\mathrm{R}_{0.8}\mathrm{Sr}_{0.2}\mathrm{NiO}_2$ ($\mathrm{R} = \mathrm{Nd}, \mathrm{La}$). At low vacancy concentrations, superconductivity is dominated by the $d$-wave pairing state. Increasing the oxygen vacancy concentration enhances the electron density of the interstitial orbital, which stabilizes an $s$-wave SC state under intermediate {\eph} coupling (i.e. $\lambda = 0.4$). This interplay yields a mixed $s+id$ pairing state. With further increasing oxygen vacancy density, the carrier density of the Ni $d_{x^2-y^2}$ orbital enters a pseudogap regime~\cite{HubbardDong2022}, rendering only pure $s$-wave superconductivity viable. Notably, this pure $s$-wave SC phase does not appear in the results of Fig.~\ref{fig:3}, which could originate from the fact that orbital occupations are governed by the bare Hamiltonian $H_0$. In realistic systems, local defects substantially redistribute the electron populations of these two orbitals. Altogether, the position dependence of the SC symmetry observed in experiments could be attributed to {\it the cooperative effects of electronic correlations, electron-phonon interactions, finite-size domains, and oxygen-defect-induced self-doping.}

In summary, we combine first-principles calculations with FLEX-ME theory to investigate the interplay between electron correlations and {\eph} coupling in IL nickelate superconductors. Our results reveal that spin fluctuations drive robust $d$-wave superconductivity on the Ni $d_{x^2-y^2}$ orbital, whereas {\eph} coupling promotes $s$-wave superconductivity in the IO band, resulting in an $s+id$ SC state. In the thermodynamic limit, the critical {\eph} coupling required to stabilize the $s+id$ state is $\lambda = 0.7$ at $n=1$. On finite-size lattices, however, this critical value can be reduced to $\lambda = 0.35$, a coupling strength that lies within the physically reasonable range estimated for $\mathrm{LaNiO}_2$.

Motivated by the multi-domain nature of realistic $\mathrm{LaNiO}_2$ samples, we further examine finite-size lattice superconductivity across various electron densities. We find that the intermediate {\eph} coupling ($\lambda = 0.4$) fails to drive $s$-wave superconductivity at $n = 0.8$ but successfully stabilizes it at $n = 0.9$. These results suggest that the spatially inhomogeneous pairing symmetry observed in $\mathrm{Nd}_{0.8}\mathrm{Sr}_{0.2}\mathrm{NiO}_2$ arises from local oxygen defects, which modulate local electron density and thereby enable distinct pairing symmetries in different domains.

Finally, we would like to emphasize that the {\eph} coupling may play a key role in enhancing the superconducting transition temperature $T_c$ under pressure~\cite{HighLee2026}. For example, Sm-based IL nickelates exhibit a $T_c$ that is twice as large as that of Nd/La/Pr-based IL nickelates~\cite{PersistentYan2026}. This elevated $T_c$ is plausibly due to the strong {\eph} coupling driven by a contracted in-plane lattice, which leads to $s$-wave superconductivity at 40~K, as shown in Fig.~\ref{fig:2}.



{\it Acknowledgement.} This work was supported by the National Natural Science Foundation of the People's Republic of China (Young Scientists Fund, Grant No. 12204236) and the Liaoning Xingliao Talent Plan. The authors also acknowledge the startup funding support from Northeastern University, Shenyang.

{\it Data Availability.} The data that support the findings of this article are not publicly available upon publication because it is not technically feasible and/or the cost of preparing, depositing, and hosting the data would be prohibitive within the terms of this research project. The data are available from the authors upon reasonable request.

\bibliography{main}

\end{document}